\documentclass[final,5p,times,twocolumn]{elsarticle}

\usepackage{amsmath}
\usepackage{graphicx}
\usepackage{dcolumn}
\usepackage{bm}
\usepackage{color}
\usepackage{bigstrut}
\usepackage{tabularx}
\usepackage{float}
\usepackage{ulem}
\usepackage[]{color}
\usepackage[colorlinks=true]{hyperref}
\newcolumntype{L}[1]{>{\raggedright\arraybackslash}p{#1}}
\newcolumntype{C}[1]{>{\centering\arraybackslash}p{#1}}
\newcolumntype{R}[1]{>{\raggedleft\arraybackslash}p{#1}}

\biboptions{square,numbers,sort&compress}

\journal{Journal of Electron Spectroscopy and Related Phenomena}

\begin{document}
\begin{frontmatter}

\title{Quantitative study of valence and configuration interaction parameters of the Kondo semiconductors Ce$M_2$Al$_{10}$ ($M$ = Ru, Os and Fe) by means of bulk-sensitive\\ hard x-ray photoelectron spectroscopy}

\address[Cologne]{Institute of Physics II, University of Cologne, Z{\"u}lpicher Stra{\ss}e 77, 50937 Cologne, Germany}
\address[Muro]{Department of Liberal Arts and Sciences, Toyama Prefectural University, Izumi 939-0398, Japan}
\address[Hiroshima1]{Department of Quantum Matter, AdSM, Hiroshima University, Higashi-Hiroshima 739-8530, Japan}
\address[Hiroshima2]{Institute for Advanced Materials Research, Hiroshima University, Higashi-Hiroshima 739-8530, Japan}
\address[NSRRC]{National Synchrotron Radiation Research Center, 101 Hsin-Ann Road, Hsinchu 30077, Taiwan}
\address[Dresden]{Max Planck Institute for Chemical Physics of Solids, N{\"o}thnizer Stra{\ss}e 40, 01187 Dresden, Germany}

\author[Cologne]{F.~Strigari\corref{label1}}
\ead{strigari@ph2.uni-koeln.de}
\author[Cologne]{M.~Sundermann}
\author[Muro]{Y.~Muro}
\author[Hiroshima1]{K.~Yutani}
\author[Hiroshima1,Hiroshima2]{T.~Takabatake}
\author[NSRRC]{K.-D.~Tsuei}
\author[NSRRC]{Y.~F. Liao}
\author[Hiroshima1]{A.~Tanaka}
\author[Dresden]{P.~Thalmeier}
\author[Dresden]{M.~W.~Haverkort}
\author[Dresden]{L.~H.~Tjeng}
\author[Cologne]{A.~Severing\corref{label2}}
\ead{severing@ph2.uni-koeln.de}

\begin{abstract}
The occupancy of the $4f^n$ contributions in the Kondo semiconductors Ce$M_2$Al$_{10}$ ($M$ = Ru, Os and Fe) has been quantitatively determined by means of bulk-sensitive hard x-ray photoelectron spectroscopy (HAXPES) on the Ce $3d$ core levels. Combining a configuration interaction scheme with full multiplet calculations allowed to accurately describe the HAXPES data despite the presence of strong plasmon excitations in the spectra. The configuration interaction parameters obtained from this analysis -- in particular the hybridization strength $V_{\textrm{eff}}$ and the effective $f$ binding energy $\Delta_{f}$ -- 
indicate a slightly stronger exchange interaction in CeOs$_2$Al$_{10}$ compared to CeRu$_2$Al$_{10}$, and a significant increase in CeFe$_2$Al$_{10}$.
This shows the existence of a substantial amount of Kondo screening in these magnetically ordered systems and places the entire Ce$M_2$Al$_{10}$ family in the region of strong exchange interactions.
\end{abstract}

\end{frontmatter}

\section{Introduction}

In cerium compounds the localized $f$ electrons of the $4f$ shell interact with the surrounding conduction electrons which leads to a screening of the localized $f$ spins (Kondo effect) as well as to an indirect Ruderman-Kittel-Kasuya-Yosida (RKKY) exchange interaction between the local spins. The latter leads to magnetic order with localized spins, the former to a non-magnetic ground state with partially delocalized $f$ electrons. The competition between these two effects governs the physics of Kondo lattice materials and is described in the Doniach phase diagram \cite{Doniach1977}.
According to this phase diagram magnetically ordered ground sates with localized $4f$ moments are expected in materials with small exchange interaction $J_\textrm{ex}$ and non-magnetic ones when $J_\textrm{ex}$ is large. A consequence of the hybridization between $f$ and conduction electrons is the opening of a hybridization gap close to the Fermi energy. In some of these compounds the Fermi energy falls into this gap so that the materials exhibit Kondo insulating, semiconducting or semimetallic behavior, depending on the gap structure (e.g$.$ CeNiSn, CeBi$_4$Pt$_3$) \cite{RISEBOROUGH_2000}. The members of the Ce$M_2$Al$_{10}$ family are classified as Kondo semiconductors with narrow, anisotropic gaps of the order of a few meV~\cite{Nishioka2009,Muro2009,Lue2010a,Muro2010,Kawamura2010,Chen2010}.

In CeFe$_2$Al$_{10}$ Kondo screening appears large and the ground state is non-magnetic \cite{Nishioka2009,Muro2009,Chen2010,Kawamura2010,Kimura2011_Ru}. However, the members with $M$\,=\,Ru and Os exhibit antiferromagnetic order at fairly high temperatures of $T_\textrm{N}=27$\,K ($M$\,=\,Ru) and $29$\,K ($M$\,=\,Os) although there are signs of a considerable amount of Kondo screening according to macroscopic and neutron measurements \cite{Strydom2009,Nishioka2009,Robert2010,Khalyavin2010,Adroja2010}.
Keeping in mind the RKKY interaction gets weaker with increasing distance between the local moments, it is amazing that magnetic order forms at all in these compounds, in which the Ce atoms are more than 5\,\AA\ apart \cite{Thiede1998,Tursina2005}, and then even at such high ordering temperatures. Following the de Gennes scaling from the Gd equivalents would imply much lower ordering temperatures \cite{Nishioka2009}. The peculiarity of the magnetic order has led to a plethora of intensive studies and made the Ce$M_2$Al$_{10}$ compounds prominent examples for systems exhibiting unconventional order \cite{Muro2009,Chen2010,Kawamura2010,Kimura2011_Ru,Kambe2010,Muro2010,Tanida2010c,Kimura2011_Os,Kondo2011,Kato2011b,Hanzawa2011b,Goraus2012,Slebarski2012,Tanida2012a,Lue2012,Kunimori2012,Robert2012,Muro2012b,Strigari2012,Strigari2013,Sera2013,Kondo2013,Adroja2013,Adroja2013_Rev,Hoshino2013,Tanida2013,Kobayashi2013,Khalyavin2013,Guo2013,Khalyavin2014,Kawabata2014,Zekko2014,Mignot2014,Ishiga2014}.

The Ce atoms are situated in a cage-like environment (space group $Cmcm$) \cite{Thiede1998,Tursina2005} and the $f$ electrons experience an orthorhombic crystal-electric field which is mainly responsible for the strong magnetic anisotropy $\chi_a>\chi_c>\chi_b$ above $T_\textrm{N}$ and to a large extent for the small ordered magnetic moments \cite{Nishioka2009,Muro2010,Tanida2010c,Muro2012b,Strigari2012,Strigari2013}. The measured moments are only slightly reduced with respect to the crystal-field-only moments \cite{Strigari2012,Strigari2013}. Moreover, spin gaps have been found by inelastic neutron scattering for $M$\,=\,Ru and Os in the ordered state (8 and 11\,meV) and for $M$\,=\,Fe in the paramagnetic state (12.5\,meV) \cite{Robert2010,Adroja2010,Robert2012,Adroja2013}. The Kondo temperatures $T_\textrm{K}$ are estimated to be 52\,K and 92\,K for $M$\,=\,Ru and Os, respectively, and beyond 300\,K for $M$\,=\,Fe \cite{Adroja2013,Adroja2013_Rev}, which in the case of the Fe compound is comparable to the expected crystal-field splitting \cite{Adroja2013}.

The mechanism of the magnetic order, with its ordered moments aligned along the $c$-axis and not the easy axis~$a$, is still an open question. There are some experimental and theoretical suggestions for the Kondo screening having an impact on the magnetic order \cite{Kondo2011,Kunimori2012}. For example, it is meant to be strongest along the $a$-direction and thus responsible for the unexpected orientation of the ordered moments \cite{Kondo2011}. This seems to be consistent with the findings of the hybridization being anisotropic \cite{Kimura2011_Os,Tanida2012a,Sera2013}. Interestingly, susceptibility measurements, muon spin relaxation and neutron diffraction show that only small amounts of electron doping with Rh or Ir -- corresponding to one extra 4$d$ or 5$d$ electron in CeRu$_2$Al$_{10}$ and CeOs$_2$Al$_{10}$, respectively -- suppress the Kondo screening and flip the ordered moments parallel to the easy axis~$a$, while the ordering temperatures remain almost unchanged \cite{Kobayashi2013,Khalyavin2013,Guo2013}. On the contrary, for light hole doping in CeOs$_2$Al$_{10}$ -- by substituting Re for Os -- the size of the ordered moments decreases significantly and their alignment along the hard axis $c$ is maintained \cite{Khalyavin2014}. A Re substitution of only 5\,\% suppresses the magnetic order completely \cite{Kawabata2014}.

It is desirable to quantify the Kondo interaction because of its apparent connection with the magnetic order. We recall that weakly hybridized Ce systems are well localized and have a valence of three (Ce$^{3+}$) and a $4f$ occupancy $n_f$\,=\,1 ($f^1$). The presence of strong hybridization leads to a partial delocalization of the $f$ electrons and the no longer integer-valent $4f$ ground state can then be written as a mixed state $|\Psi_{\textrm{GS}} \rangle = \alpha$~$|f^0\rangle$~+~$\beta$~$|f^1$\underline{L}$\rangle$~+~$\gamma$~$|f^2$\underline{\underline{L}}$\rangle$ with additional contributions of the divalent and tetravalent states ($f^2$ and $f^0$). Here \underline{L} and \underline{\underline{L}} denote the number of ligand holes. The amount of $f^0$ quantifies the degree of delocalization, which in the case of a moderately large $\alpha^2$ is a synonym for the effectiveness of the Kondo screening. Core level spectroscopy techniques, like x-ray absorption or photoelectron spectroscopy (PES), are capable of seeing the different valence states because they involve the presence of a core-hole with an attractive potential which acts differently on the different $f^n$ states. As a result the valence states are re-ordered energetically: $|\underline{c}f^2\underline{\underline{L}}\rangle$ becomes the lowest configuration (with \underline{c} denoting the core hole), followed by $|\underline{c}f^1\underline{L}\rangle$ (typically $\Delta E_{f^1f^2} \approx 5$\,eV) and $|\underline{c}f^0\rangle$ (typically $\Delta E_{f^0f^1} \approx 11$\,eV), yielding three spectral features which correspond to the final states with mainly \underline{c}$f^2$\underline{\underline{L}}, \underline{c}$f^1$\underline{L} and \underline{c}$f^0$ character \cite{Gunnarsson2001}.    
The corresponding spectral intensities $I(\underline{c}f^0)$, $I(\underline{c}f^1\underline{L})$ and $I(\underline{c}f^2\underline{\underline{L}})$ contain information about $\alpha^2$, $\beta^2$ and $\gamma^2$, respectively, so that the $f$ electron count $n_f=\beta^2+2\gamma^2$ can be deduced. Note, because of hybridization effects in the final state, the spectral intensities are not directly proportional to $\alpha^2$, $\beta^2$ and $\gamma^2$. The \textsl{translation} of $I(\underline{c}f^n)$ to the actual $f^n$ contributions $\alpha^2$, $\beta^2$ and $\gamma^2$ in the ground state is achieved using the full multiplet configuration interaction calculations as explained below. In the following we will use the short notation $f^0$, $f^1$ and $f^2$ only, omitting the explicit notation of the core and ligand holes for simplicity.

We have carried out hard x-ray photoelectron spectro\-scopy \mbox{(HAXPES)} on the Ce $3d$ core levels of Ce$M_2$Al$_{10}$ in order to determine the occupancy of the $4f$ shell in Ce$M_2$Al$_{10}$. Soft x-ray PES has proven to be a valuable technique for the investigation of the electronic states of rare earth compounds \cite{Allen1986,Huefner1992,Tjeng1993}, but suffers from surface effects. Especially in correlated electron systems the degree of hybridization at the surface is known to be reduced with respect to the bulk \cite{Laubschat1990,Suga_Ce_2000,Suga_Sm_2007}. The use of hard x-rays provides the bulk sensitivity needed to image the bulk electronic structure in these systems \cite{Suga_Sm_2007,Braicovich1997,TanumaIMFP_2011}.

\section{Experimental details}

Polycrystals of CeRu$_2$Al$_{10}$, CeOs$_2$Al$_{10}$ and CeFe$_2$Al$_{10}$ were synthesized by arc melting under an argon atmosphere and the sample quality and stoichiometry were confirmed by powder x-ray diffraction and electron-probe microanalysis \cite{Muro2009}. The HAXPES measurements were performed at the Taiwan beamline BL12XU at SPring-8, Japan, with an incident photon energy of 6.47\,keV and at an incidence angle of $45^{\circ}$. For the determination of $E_\textrm{F}$ the valence band spectrum of a Au film was measured. The excited photoelectrons were collected and analyzed (MB~Scientific A-1 HE) in the horizontal plane at an emission angle of $45^{\circ}$ in ultrahigh vacuum with a base pressure of $10^{-9}$\,mbar. Clean sample surfaces were obtained by cleaving the polycrystals \textsl{in situ} at low temperature ($T\leq60$\,K). Multiple single scans were recorded over a time period of several hours. Their reproducibility ensured that clean surfaces were maintained over time. The overall energy resolution was about 1\,eV in the energy region of the Ce\,$3d$ emission.

\section{Experimental results}
\label{ExpResults}

\begin{figure}[!t]
   \includegraphics[width=0.995\columnwidth]{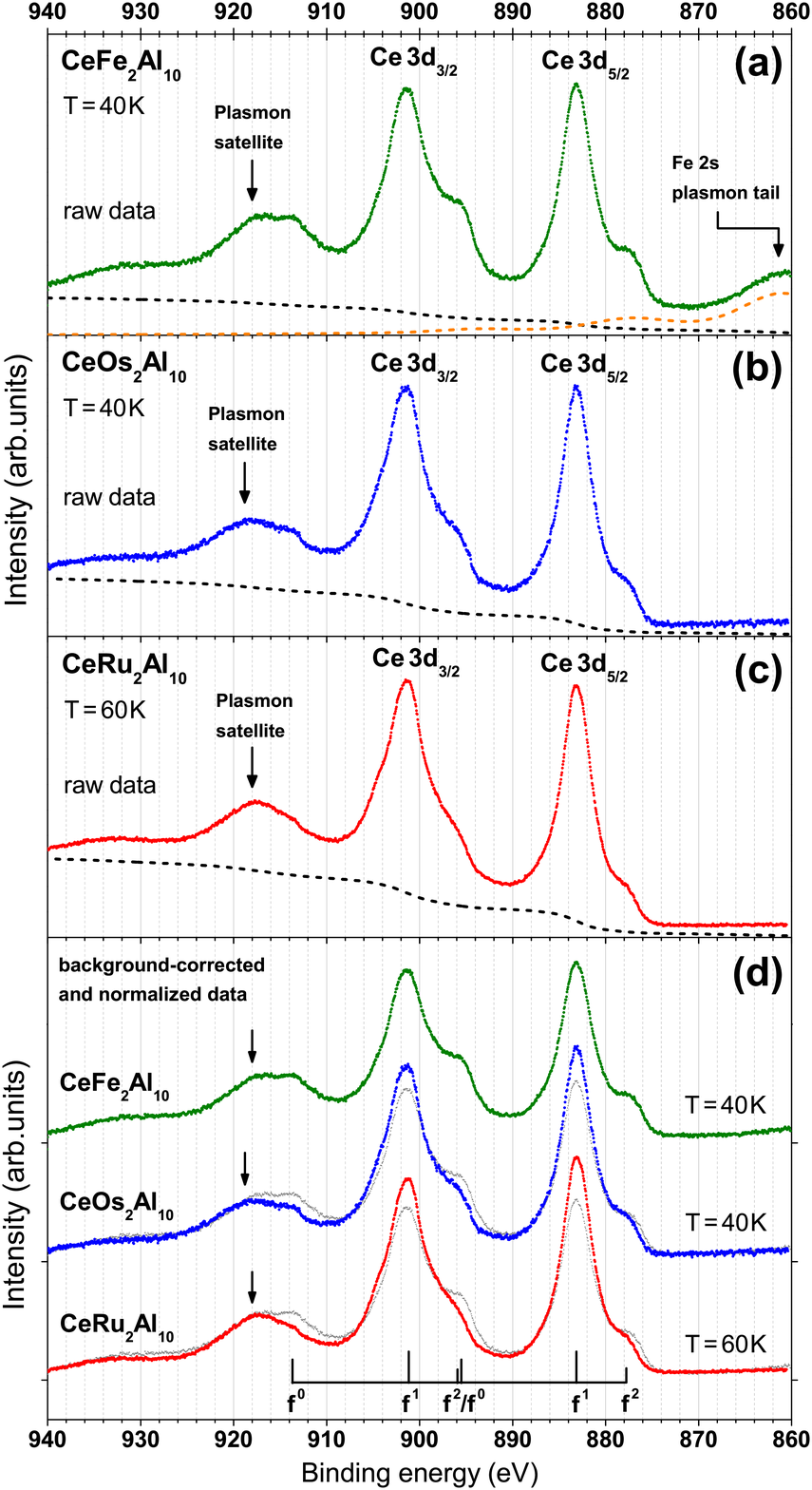}
    \caption{Low temperature Ce~$3d$ HAXPES spectra of Ce$M_2$Al$_{10}$ with $M$ = Ru (red), Os (blue), and Fe (green). (a)-(c) Data as measured, i.e$.$ before background subtraction and normalization. The black arrows at $\sim\!918$\,eV indicate spectral weight due to plasmon contributions. In CeFe$_2$Al$_{10}$ another plasmon, originating from the Fe~$2s$ emission, shows up. Its estimated contribution is plotted as orange dashed line in (a). The black dashed line in each panel shows the respective integral background. (d) Data after background subtraction and normalization. All data are normalized to the integrated intensity between 870 and 940\,eV and the three spectra are displaced on the y-axis by an offset. For a better comparison, the CeFe$_2$Al$_{10}$ spectrum is additionally overlayed on the Ru and Os data (in gray). The black ruler at the bottom indicates the energy positions of the $f^n$ contributions and the black arrows indicate spectral weight due to plasmon contributions.}
    \label{rawdata}
\end{figure}

In Fig.\,\ref{rawdata} the Ce~$3d$ core level HAXPES spectra of Ce$M_2$Al$_{10}$ with $M$ = Ru (red), Os (blue), and Fe (green) are shown. For simplicity the different compounds are referred to as \textsl{Ru}, \textsl{Os} and \textsl{Fe} in the following. The measurements were carried out at 40\,K for Fe and Os and at 60\,K for Ru which is low enough to be in the Kondo regime. In the panels (a)\hbox{-}(c) the raw data are shown. All spectra exhibit very low statistical noise and were highly reproducible. The dashed black lines display the standard integral background as developed by Shirley \cite{Shirley1972}.

The main emission lines at about 883 and 901.5\,eV binding energy represent the predominant spin-orbit split Ce~$3d_{5/2}\,f^1$ and $3d_{3/2}\,f^1$ multiplets. The mixed ground state character is reflected in additional spectral weight at the shoulders of the $f^1$ structures ($f^2$ contributions, $\sim$\,5.5\,eV on the lower binding energy side). The $3d_{3/2}\,f^0$ feature comes up at $\sim$\,914\,eV binding energy, whereas the $3d_{5/2}\,f^0$ largely overlaps with the $3d_{3/2}\,f^2$ features at about 895\,eV. Strikingly, additional broad humps show up in all spectra at about 918\,eV and in the Fe sample also around 860\,eV (indicated by black arrows in Fig.\,\ref{rawdata}). These humps are identified as plasmon resonances \cite{Huefner}.

As typical for cage-like structures, in the case of Ce$M_2$Al$_{10}$ the plasmon excitations originate from the polyhedral aluminum cage surrounding the Ce atom. Plasmon peaks of first and higher orders appear for each emission and multiplet line at a fixed energy distance. The main plasmonic contributions -- which become notably visible at about 918\,eV (see black arrows in Fig.\,\ref{rawdata}) -- belong to the Ce~$3d$ emission and have unfortunately a large overlap with the $3d_{3/2}\,f^0$ feature, thereby preventing a direct extraction of its spectral weight.
In CeFe$_2$Al$_{10}$ the Fe~$2s$ emission at $\sim$\,845\,eV gives rise to an additional plasmon satellite peaking at 860\,eV. Its estimated contribution (up to third order) is shown in panel~(a) of Fig.\,\ref{rawdata} as dashed orange line. Three Voigt profiles are used here and their intensity and line width is estimated on the basis of the Al~$1s$ analysis (see \ref{AppendixA}). Overcoming the drawbacks due to the appearance of plasmons in the Ce~$3d$ HAXPES spectra has been an important task for the data analysis.

In Fig.\,\ref{rawdata}(d) the background-corrected and normalized Ce~$3d$ core level HAXPES spectra of Ce$M_2$Al$_{10}$ with $M$~=~Ru (red), Os (blue), and Fe (green) are shown. The integral backgrounds displayed in Fig\,\ref{rawdata}(a)-(c) have been subtracted from the data and the spectra have been normalized to the integrated intensity between 870 and 940\,eV. In the case of Fe the estimated contribution from the plasmon satellites belonging to the Fe~$2s$ emission has also been subtracted. The energy positions of the different $f$ contributions are indicated by the black ruler at the bottom of Fig.\,\ref{rawdata}(d). Additionally, for a better visualization of the spectral differences between the three compounds, the CeFe$_2$Al$_{10}$ spectrum is overlayed (gray curves) on the Ru and Os data.
Comparing the three spectra with each other reveals already a qualitative trend for the $f$ occupancy: While the $f^1$ contributions to the spectrum decrease from Ru to Os to Fe, the $f^0$ (tiny spikes on top of the plasmon intensities) and $f^2$ spectral weights become more pronounced in the same direction. This points towards an increasing $f$ delocalization in the same order, in agreement with previous experimental findings \cite{Nishioka2009,Muro2010,Kimura2011_Ru,Goraus2012,Slebarski2012,Adroja2013,Strigari2013,Zekko2014}. However, a \textsl{quantitative} extraction of the different $f^n$ contributions to the HAXPES spectrum requires an adequate modeling of the plasmon contributions arising from the Ce~$3d$ emission lines.

\section{Quantitative analysis}
\label{QuantAnalysis}

\subsection{Concept}
Anderson proposed an impurity model to explain the moment of magnetic impurities in nonmagnetic host metals \cite{Anderson1961}. The model was extended to the analysis of x-ray absorption and PES spectra of mixed valence Ce compounds by Gunnarsson and Sch\"onhammer \cite{Gunnarsson1983}. It considers a single $f$ state in a bath of electrons, which are described in a band model, and the hybridization between them. Examples for its successful application to PES, core level spectroscopy and x-ray absorption data can be found in Refs.~\cite{Allen1986,KotaniJo1988,Laubschat1990,Huefner1992,Tjeng1993}. A more recent example are the $3d$ core level HAXPES data of CeRu$_2$Si$_2$ and CeRu$_2$Ge$_2$ by Yano~\textsl{et al.} \cite{Suga2008} However, these descriptions do not include a multiplet calculation, despite the complex underlying multiplet structure resulting from the $f$\hbox{-}$f$~Coulomb and exchange interactions, because computing times would become unreasonably long.

For our case here we do need to include a full multiplet calculation because the line shapes in the Ce~$3d$ emission spectra, which are primarily determined by the underlying multiplet structure, are complicated by the strong plasmons. Each emission line of the two spin-orbit split $3d$ multiplets gives rise to first and higher order plasmons, thus preventing a simple phenomenological assignment of the respective $f^n$ spectral weights with Gaussian and/or Lorentzian line profiles. In order to tackle this problem we had to combine a full multiplet calculation with a simpler form of the Anderson impurity model; simpler in order to keep computing times reasonable. This simplified form is a configuration interaction (CI) calculation in which the valence band is represented by a single ligand state. It captures the fundamental features of the core hole spectrum \cite{Imer1987}, i.e. it yields accurate $f^n$ contributions and the resulting CI parameters give insight into the exchange interaction $J_{ex}$. The consequences of the simplification are discussed in \ref{AppendixB}.

Plasmons appear at well-defined energy distances at higher binding energies ($\Delta E_\textrm{n} = \textrm{n} E_{\textrm{plasmon}}$) and the application of the full multiplet calculation allows the pinning of a plasmon and its multiples to each emission line with the same parameters for energy distance, line width and shape. The line shape parameters for the plasmon satellites can be determined from the Al~$1s$ single emission line in an independent measurement (see \ref{AppendixA} and Fig.~\ref{plasmon} therein), reducing the number of free fit parameters for the \textsl{reconstruction} of the spectra (Ce $3d$ plus plasmons intensities). The combination of full multiplet and CI calculations (fm-CI), allows to extract the $f^n$ contributions despite strong plasmons.

The fm-CI simulations were performed with the XTLS 9.0 program \cite{TanakaJPSC63}. They account for the intra-atomic $4f$-$4f$ and $3d$-$4f$ Coulomb and exchange interactions and the $3d$ and $4f$ spin-orbit coupling, as calculated with Cowan's atomic structure code \cite{Cowan}. From earlier studies \cite{Strigari2012,Strigari2013}, the reduction of the atomic Hartree-Fock values for the $4f$-$4f$ and $3d$-$4f$ Coulomb interactions are known to amount to $\sim$\,40\,\% and $\sim$\,20\,\%, respectively. The hybridization effects between the $f$ and the conduction electrons are described by the $f$-$f$ Coulomb exchange ($U_{ff}$), the Coulomb interaction between $f$ electron and $d$ core hole ($U_{fc}$), the effective $f$ binding energy $\Delta_{f}$ (i.e. the energy difference between $f^0$ and $f^1$\b{L} in the initial state) and the hybridization strength $V_{\textrm{eff}}$. Thus, in total there are four parameters plus line shape to be fitted (i.e$.$ Lorentzian and Mahan broadening, see below). The energy distances between the $f^n$ features and their respective intensities uniquely determine the four CI parameters. The plasmon line shape and properties are fixed by the independent analysis of the Al~$1s$ measurements.

\subsection{\texorpdfstring{Simulation of Ce~$3d$ spectra}{Simulation of Ce 3d spectra}}
For each Ce~$3d$ multiplet line, plasmonic contributions up to the order of n\,=\,3 are included using the line shape parameters and intensity ratios as determined in the analysis of the Al~$1s$ spectra (see \ref{AppendixA}). Thus, each \textsl{line profile} consists of the main emission lines plus first, second and third order plasmon. The same line shape and intensity ratios were used for all $3d$ emission lines. Having fixed the line profiles, the spectra were calculated using the fm-CI routine.

\begin{figure}[!t]
   \includegraphics[width=0.98\columnwidth]{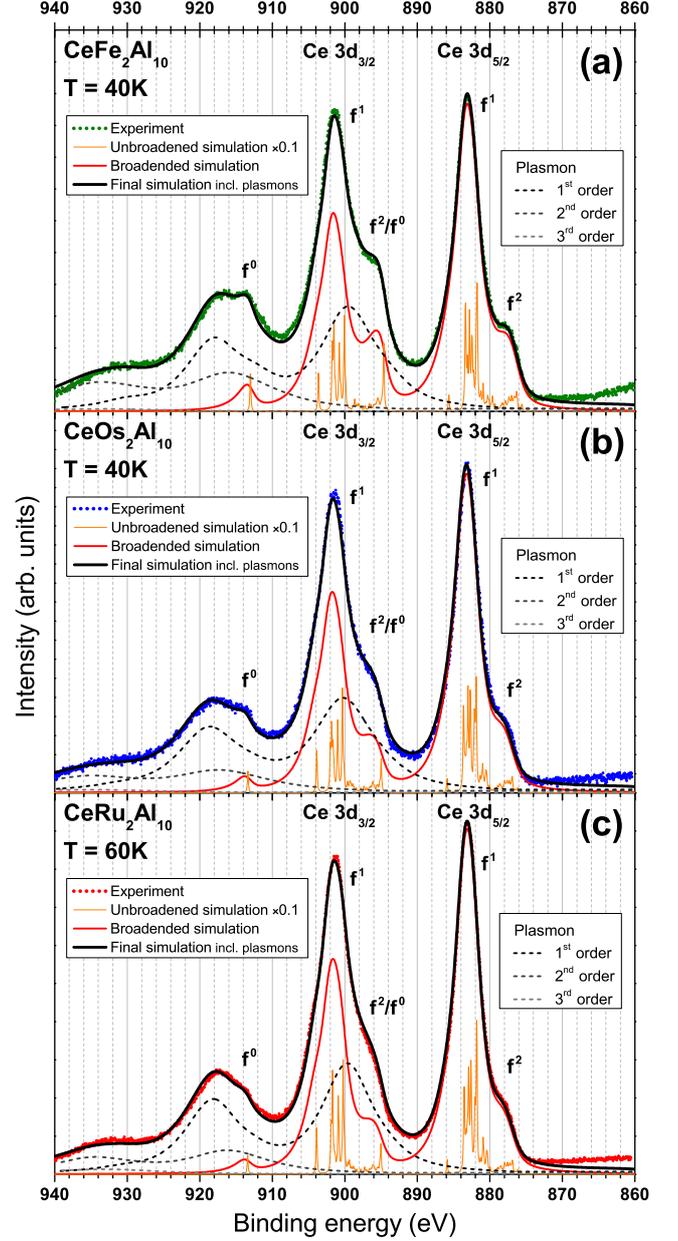}
    \caption{Transfer of the Al plasmon properties -- as extracted from the Al $1s$ spectra -- to the fm-CI simulation of the Ce $3d$ core emission for (a) CeRu$_2$Al$_{10}$, (b) CeOs$_2$Al$_{10}$ and (c) CeFe$_2$Al$_{10}$. The experimental data is shown as green, blue and red dots, respectively. (Same as in Fig.~\ref{rawdata}) The simulated multiplet structure is shown as orange (unbroadened, scaled by $\times0.1$) and red (broadened) line curves. The calculated spectral weight due to plasmons (up to the third order) is represented by the dashed lines. The final simulations are represented by the black curves.}
    \label{plasmon_corr}
\end{figure}

This procedure is visualized in Fig.\,\ref{plasmon_corr} (a), (b) and (c) for Fe, Os and Ru. In each panel the dots represent the low temperature Ce~$3d$ HAXPES data (same color code as in Fig.\,\ref{rawdata}).
The multiplet simulation (orange curve) is broadened by a Gaussian function of 1\,eV FWHM and a Lorentzian function of 1.3\,eV FWHM to account for the instrumental resolution and lifetime broadening, respectively. In addition a Mahan function (cut-off parameter $\xi=1.8$\,eV, asymmetry factor $\alpha=0.65$) is used to account for the asymmetric line shape of the $3d$ emission (see Eq.~(1) in \ref{AppendixA}). The red curves in Fig.\,\ref{plasmon_corr} are the resulting broadened Ce $3d$ multiplet spectra. The dashed lines represent the plasmons (first (black), second (dark gray) and third order (gray)) which are replica of the red curve using the intensity scaling factors, energy shifts and broadening as obtained from the fits to the Al~$1s$ spectra (see \ref{AppendixA}, Table~\ref{plasmonfit}). The black lines are the total fit to the data, i.e$.$ the sum of the red curve ($3d$ emission) and all dashed lines (plasmons). Note that a change of the configuration interaction parameters in the theoretical model not only changes the multiplet structure but also leads to different plasmon intensity contributions.

In the top rows of Table~\ref{CIparam} the resulting $f$ weights and the $f$ electron count $n_f$ are given. The corresponding fit parameters for the $f$-$f$ Coulomb exchange ($U_{ff}$), the Coulomb interaction between $f$ electron and $3d$ core hole ($U_{fc}$), the hybridization strength $V_{\textrm{eff}}$ and the effective $f$ binding energy $\Delta_{f}$ (all given in eV) are listed in the bottom rows. These results describe the pure, background and plasmon corrected Ce~$3d$ HAXPES spectra (red line curves in Fig.\,\ref{plasmon_corr}).

\begin{table}[tbp]
  \centering
    \begin{tabular*}{0.48\textwidth}{@{\extracolsep{\fill}}rrrr}
    \hline
		\hline
          & \multicolumn{1}{c}{CeRu$_2$Al$_{10}$} & \multicolumn{1}{c}{CeOs$_2$Al$_{10}$} & \multicolumn{1}{c}{CeFe$_2$Al$_{10}$} \bigstrut[m]\\
    \hline
		\multicolumn{1}{l}{$f^0$} & 5.9(8)\,\%  & 7.2(8)\,\%  & 11.6(8)\,\% \bigstrut\\
    \multicolumn{1}{l}{$f^1$} & 91.7(8)\,\% & 90.1(8)\,\% & 85.0(8)\,\% \bigstrut\\
    \multicolumn{1}{l}{$f^2$} & 2.4(4)\,\%  & 2.7(4)\,\%  & 3.5(4)\,\% \bigstrut\\
    \multicolumn{1}{l}{$n_f$} & 0.97(1) & 0.96(1) & 0.92(1) \bigstrut\\
		\hline
		\multicolumn{1}{l}{$U_{ff}$} & 8.6(2)  & 8.5(2)   & 8.0(2) \bigstrut\\
    \multicolumn{1}{l}{$U_{fc}$} & 10.00(15) & 10.00(15) & 10.30(15) \bigstrut\\  
    \multicolumn{1}{l}{$V_{\textrm{eff}}$} & 0.214(5) & 0.230(5) & 0.268(5) \bigstrut\\
    \multicolumn{1}{l}{$\Delta_{f}$} & -2.4(1)  & -2.3(1)  & -1.9(1) \bigstrut\\
    \hline
		\hline
    \end{tabular*}
  \caption{Results from fitting the background- and plasmon-corrected Ce~$3d$ HAXPES spectra of the Ce$M_2$Al$_{10}$ compounds within the fm-CI model (see Fig.\,\ref{plasmon_corr} for the corresponding simulations). In the top rows the determined $f^n$ contributions and $f$ electron count $n_f$ are given. The corresponding CI parameters are listed in the bottom rows: the $f$-$f$ Coulomb exchange $U_{ff}$, the Coulomb interaction between $f$ electron and $3d$ core hole $U_{fc}$, the effective $f$ binding energy $\Delta_{f}$, and the hybridization strength $V_{\textrm{eff}}$ (all in eV).}
  \label{CIparam}
\end{table}

\section{Discussion}
\label{Disc}
There is an excellent agreement between the theoretical and the experimental spectra for all three compounds (see Fig.~\ref{plasmon_corr}). All spectral features -- including the different $f^n$ contributions -- and the plasmon humps are very well reproduced. The agreement is almost perfect for the Ce~$3d_{5/2}$ emission lines since the corresponding energy region is hardly influenced by plasmons and the background contribution can be assumed to be well determined. On the other hand, also above 892\,eV binding energy the general Ce~$3d_{3/2}$ line shape and the $f^0$ and $f^2$ spectral features are nicely modeled by the fm-CI calculation. In particular, the $f^0$ feature can be extracted reliably because it consists of only one single emission line and, therefore, appears very narrow compared to background and plasmons.

The overall size of the resulting configuration interaction parameters (see Table~\ref{CIparam}) is comparable with the results of e.g$.$ the HAXPES analysis of CeRu$_2$Si$_2$ by Yano \textsl{et al.} \cite{Suga2008} in which the Anderson impurity model is used without considering multiplet effects. In recent resonant PES measurements on Ce$M_2$Al$_{10}$ the energy separation of the $f^0$ feature is found to be approximately 2\,eV \cite{Ishiga2014}, which is consistent with our results for $\left|\Delta_{f}\right|$.
The trend observed for the $4f$ valence in Ref.\,\cite{Ishiga2014} as well as in earlier core level x-ray photoemission studies \cite{Kimura2011_Ru,Goraus2012,Slebarski2012} is also generally in agreement with our findings, although the deviations from integer valence are smaller. It should be noted that these studies do not reach the same bulk sensitivity due to the use of soft x-rays \cite{Suga_Ce_2000,Suga_Sm_2007,Braicovich1997,TanumaIMFP_2011}. The probing depth for Ce~$3d$ core-level PES given by the photoelectrons' inelastic mean free path at $h\nu = 6.47$\,keV is about 85\,\AA, whereas for soft x-rays ($h\nu = 1000 - 1500$\,eV) it amounts to $5 - 15$\,\AA.  Zekko \textsl{et al$.$} \cite{Zekko2014} investigated the substitution series Ce(Ru$_{1-x}$Fe$_x$)$_2$Al$_{10}$ by means of bulk-sensitive partial fluorescence yield x-ray absorption (PFY-XAS) at the Ce~$L_3$ edge and also found a more pronounced difference between the Ce valence of CeRu$_2$Al$_{10}$ ($n_f=0.95$) and CeFe$_2$Al$_{10}$ ($n_f=0.89$).

The question arises to what extent the difference of surface sensitivity between HAXPES and PFY-XAS has an impact on the extraction of the $f^n$ contributions. The latter is a photon-in-photon-out technique and thereby provides a probing depth of the order of 10\,$\mu$m, making PFY-XAS truly bulk-sensitive. The bulk sensitivity of high-energy Ce~$3d$ PES has been studied in detail by Braicovich \textsl{et al.} \cite{Braicovich1997}, showing that for $h\nu = 3.85$\,keV the bulk contribution to the spectra already dominates. Hence, in the present Ce$M_2$Al$_{10}$ spectra for $h\nu = 6.47$\,keV the remaining influence from the surface region is estimated to be non-relevant, especially because the samples were cleaved \textsl{in situ} under ultrahigh vacuum conditions and surface degradation with time was not observed.

While we believe that PFY-XAS is very powerful in determining small changes in the $f^n$ contributions with temperature and especially with pressure, we also believe that there are non-negligible ambiguities in the absolute assignment of the spectral weights. PFY at the Ce $L$-edge measures the empty $5d$ density of states so that band structure determines the spectral distribution, giving rise to non-trivial line shapes. For example, most of the fits require a not further specified so-called satellite peak on the high energy side of the main $f^1$ feature \cite{Rueff_CeCu2Si2,Yamaoka_CePd2Si2,Zekko2014}. Moreover, different groups place the $L_3$ edge jump at different energies (cf$.$ Ref.~\cite{Zekko2014} and \cite{Rueff_CeCu2Si2}) which may give rise to an absolute difference of a few percent in $n_f$. Band structure and lattice effects would have to be considered to describe the spectral shape and background, and be able to extract absolute values for the $f^n$ contributions, in particular to quantify the $f^0$ weight.

Back to the present HAXPES study, the analysis within the combined fm-CI model and the accurate treatment of the plasmon excitations gives a highly quantitative picture of the $4f$ ground state in the Ce$M_2$Al$_{10}$ compounds. The results indicate a substantial delocalization of the $f$ electrons in all three compounds of the Ce$M_2$Al$_{10}$ family -- being slightly stronger in Os with respect to Ru and considerably larger in Fe. $V_\textrm{eff}$ increases and $\left|\Delta_{f}\right|$ decreases from Ru to Os to Fe, meaning that $J_\textrm{ex}$ is a little bit larger for Os compared to Ru and significantly greater for Fe. Here we assume $J_\textrm{ex}$ scales with the inverse of $\left|\Delta_{f}\right|$ and quadratically with $V_\textrm{eff}$. The increasing exchange interaction is consistent with the simultaneous increase of $f^0$ and $T_\textrm{K}$ when going from Ru to Os to Fe.
In total we conclude that $J_\textrm{ex}$ is large. Consequently, due to the resulting moment screening the Ce$M_2$Al$_{10}$ family has to be considered as a \textsl{correlated} material and the de Gennes scaling law has lost its validity to predict ordering temperatures. Intuitively, in the magnetic region of the Doniach phase diagram a large $J_\textrm{ex}$ should lead to high ordering temperatures when the $1/R^3$ dependent RKKY interaction has a large amplitude. However, the distance between the local moments in these compounds are very large so that the ordering temperatures can only be high, when the RKKY interaction is very effective. This points to band structure effects which are not contained in the Doniach model. Furthermore, while the absence of magnetic order in the Fe compound can be understood within the Doniach model, it is not possible to explain its existence in the presence of a substantial amount of delocalized $f$ electrons in the Ru and Os compounds. Both have as much $f^0$ in the ground state ($\approx\!6$\,\%) as the non-magnetic compound CeRu$_2$Si$_2$ \cite{Suga2008}.

The importance of the band structure is supported by the detailed macroscopic investigation of the CeOs$_2$Al$_{10}$ substitution series where Os has been substituted with Re and Ir \cite{Kawabata2014}. Here hole doping (Re) increases the hybridization while electron doping (Ir) leads to a stronger $f$ electron localization. The important finding is that, as a function of substitution, the maximum in $T_\textrm{N}$ coincides with the maximum in the hybridization gap, thus pointing towards a connection between the two \cite{Kawabata2014}. Further studies are on their way.

The coexistence of magnetic order and $f$ delocalization has been discussed in the context of the CeRh$_{1-x}$Co$_x$In$_5$ substitution series where the $f$ electrons, with increasing Co content, change from fairly localized to more itinerant well inside the antiferromagnetic phase \cite{Ce115_Goh2008}. The Doniach phase diagram cannot capture this. When it comes to a theoretical description, the original continuum Anderson single impurity model treats the Kondo aspect and a two-impurity model the RKKY interactions -- but in reality we are dealing with Kondo lattices. There have been some models which go beyond the Anderson impurity ansatz and yield both on-site Kondo and inter-site RKKY correlations. For example Zerec \textsl{et al.} \cite{Zerec_2006} use the Kondo lattice model (KLM) ansatz for Kondo clusters and their results show the importance of the electron density of states for the competition of Kondo screening and RKKY interactions. Another model treating the two-impurity model analytically points out the importance of the local moment separation \cite{Santoro_1994}. Recently, Hoshino and Kuramoto suggested an extended phase diagram based on the KLM in a simplified approach \cite{Hoshino2013}. Here the on-site Kondo screening is obtained in a dynamical mean-field theory (DMFT) and the RKKY interaction is added in a molecular field approximation to investigate how the mean-field magnetic states are influenced by the Kondo interaction. The model yields regions of coexisting RKKY interaction and Kondo screening for large exchange interactions and the authors point out that the Ru and Os compounds of the Ce$M_2$Al$_{10}$ family fall into this region of coexistence, while the Fe compound is located in a region where magnetic order is suppressed due to the strong Kondo effect. However, none of the above mentioned theories is able to make quantitative predictions and we conclude that it would be highly desirable to have theories which also include explicitly hybridization gaps as well as crystal-field effects to account for possible anisotropies in the hybridization.

\section{Summary}
We have presented bulk-sensitive Ce~$3d$ HAXPES data of the Kondo semiconducting Ce$M_2$Al$_{10}$ family and shown that a quantitative analysis of the Ce $4f$ valence is possible despite strong plasmonic contributions in the spectra. The data were analyzed using a full multiplet configuration interaction model in which a single ligand state is used to mimic the valence band. The impact of this simplification is discussed in the Appendix.
On the basis of the full multiplet structure the line shapes of the $f^1$ and $f^2$ final states could be well described and the spectral background and the broad plasmon satellites consistently modeled, so that also the narrow $f^0$ feature was quantified, showing that strong $f^0$ contributions are present in the spectra of all three compounds of the Ce$M_2$Al$_{10}$ family. The deduced configuration interaction parameters clearly indicate an increasing trend for $J_\textrm{ex}$ from Ru to Os to Fe and in general $J_\textrm{ex}$ is concluded to be large, showing the existence of $f$ delocalization (Kondo screening) in the presence of magnetic order (RKKY interaction).\\

This work was supported by Deutsche Forschungsgemeinschaft, Germany (Grant No. 583872), and a Grant-in-Aid of MEXT, Japan
(Grant No. 26400363).

\appendix
\section{Plasmon properties}
\label{AppendixA}

\begin{table*}
  \centering
  \renewcommand{\arraystretch}{0.85}
  \begin{tabular*}{0.98\textwidth}{@{\extracolsep{\fill}}rrrrrrrrrr}
  	\hline
		\hline
          & \multicolumn{3}{c}{CeRu$_2$Al$_{10}$} & \multicolumn{3}{c}{CeOs$_2$Al$_{10}$} & \multicolumn{3}{c}{CeFe$_2$Al$_{10}$} \bigstrut[m]\\
    Order of plasmon n & \multicolumn{1}{r}{1} & \multicolumn{1}{r}{2} & \multicolumn{1}{r}{3} & \multicolumn{1}{r}{1} & \multicolumn{1}{r}{2} & \multicolumn{1}{r}{3} & \multicolumn{1}{r}{1} & \multicolumn{1}{r}{2} & \multicolumn{1}{r}{3} \bigstrut\\
		\hline
    Scaling factor & 0.65  & 0.35  & 0.12   & 0.63  & 0.33  & 0.10  & 0.68  & 0.40  & 0.12 \bigstrut\\ 
    Energy shift $\Delta E_\textrm{n}$ (eV) & 16.6 & 33.2  & 49.8 & 17.2  & 34.4  & 51.6  & 16.5  & 33.0    & 49.5 \bigstrut\\    
    Lorentzian FWHM (eV) & 5   & 10     & 15   & 6     & 12     & 18     & 5   & 10     & 15 \bigstrut\\
    \hline
		\hline
    \end{tabular*}
  \caption{Summary of the properties of the Al plasmons in Ce$M_2$Al$_{10}$ as obtained from fitting the Al~$1s$ HAXPES spectra. The scaling factor gives the plasmon intensity with respect to the main $1s$ emission line, $\Delta E_\textrm{n}$ is the energy distance relative to the position of the main line and in the last row the applied additional Lorentz broadening is noted.}
  \label{plasmonfit}
\end{table*}

\begin{figure}
   \includegraphics[width=0.99\columnwidth]{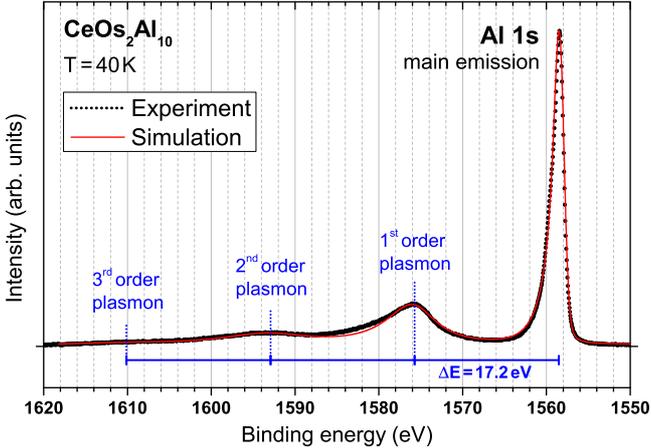}
    \caption{Low temperature HAXPES spectrum of the Al~$1s$ core level and its plasmon satellites, measured on CeOs$_2$Al$_{10}$. The black dots show the experimental data and the red line represents the simulation of the $1s$ emission including the plasmon contributions up to the third order (see text). The blue ruler at the bottom marks the energy position of the plasmon peaks relative to the main line.}
    \label{plasmon}
\end{figure}

Here the determination of the plasmon properties from the Al~$1s$ HAXPES data shall be described.
As mentioned in Sec.\,\ref{ExpResults}, the photoemission process often gives rise to plasmon satellites of first and higher order. The properties of these secondary excitations relative to the respective main emission line (i.e$.$ intensity, energy distance and broadening) are the same for each emission. In order to extract the characteristic plasmon properties and thereby be able to model the plasmon contributions in the Ce~$3d$ spectra quantitatively, we measured the Al~$1s$ emission for each compound in the energy region from 1550\,eV to 1620\,eV . The Al~$1s$ core level gives rise to a single well defined emission line without overlap with other core levels, thus is easy to model. Fig.\,\ref{plasmon} shows the background-corrected HAXPES Al~$1s$ spectrum measured for CeOs$_2$Al$_{10}$ (black dots) at 40\,K. For the background correction the standard integral background was used \cite{Shirley1972}. Apart from the main emission at 1558.5\,eV the spectrum shows at least two plasmon satellites (first and second order) on the higher binding energy side. At around 1610\,eV a third order plasmon is faintly visible. The $1s$ feature exhibits a slight asymmetric line shape, characteristic for metals \cite{Huefner} -- which is also observable in the first plasmon peak -- with a tail extending to higher binding energies. To simulate the spectrum the discrete emission line is convoluted with a Gaussian and a Lorentzian function of 1\,eV and 0.2\,eV FWHM, respectively. The so-called Mahan broadening function
	\begin{equation}
			B_{\mathrm{Mahan}}(\omega)=\frac{\Theta(\omega)}{\Gamma(\alpha) \cdot \omega}\left(\frac{\omega}{\xi}\right)^\alpha e^{-\omega/\xi}
	\end{equation}
describes the asymmetry. Here $\Theta(\omega)$ is the Heaviside step function and $\Gamma(\alpha)$ the Gamma function. The cut-off parameter is $\xi =1.8$\,eV and the asymmetry factor is about $\alpha=0.5$ for all three compounds.

In a second step the plasmons are fitted: Each plasmon of the order n is generated by duplicating the main emission line, shifting it in energy ($\Delta E_\textrm{n}=\textrm{n}E_{\textrm{plasmon}}$) and adding it to the theoretical spectrum. The plasmon features are additionally broadened by a Lorentzian and scaled in intensity to match the experimental data. The broadening applied to the first order is doubled for the second order and tripled for the third order plasmon.
The red line in Fig.\,\ref{plasmon} represents the fitting result for CeOs$_2$Al$_{10}$ containing the Al~$1s$ emission and plasmon contributions up to the third order. The agreement between experiment and simulation is very good.
The Al~$1s$ spectra of CeRu$_2$Al$_{10}$ and CeFe$_2$Al$_{10}$ have been analyzed correspondingly (not shown here), yielding an independent and complete characterization of the plasmon features for each sample. All fitting results are summarized in Table~\ref{plasmonfit}. The plasmon attributes obtained from the Al $1s$ spectra of the three compounds show a high resemblance, nevertheless there are faint but notable differences -- especially regarding the energy positions (see Table~\ref{plasmonfit}).

\section{Consequences of the fm-CI model}
\label{AppendixB}
Here the consequences of using the fm-CI model, which has been applied for the analysis of the present HAXPES data, shall be discussed. The essential simplification is the representation of the valence states by one ligand state, i.e$.$ by an infinitely narrow band. This model reproduces the main features of the core level spectra (see e.g$.$ Ref.\,\cite{Imer1987}) and has the great advantage that the computational aspect becomes easy to handle, so that it can be combined with a full multiplet calculation. However, it fails to account for the low-energy excitations, as e.g$.$ in valence band PES, and does not give realistic values for the Kondo temperature and RKKY interaction.

Another artifact of the fm-CI is the overestimation of the contribution of the $J=7/2$ multiplet in the ground state. Here we recall that in cerium the spin-orbit splitting ($\Delta_\textrm{SO} \approx 0.3$\,eV) is often of the order of the hybridization $V_{\textrm{eff}}$ (see Table~\ref{CIparam}) so that the higher multiplet intermixes with the ground state, the more the stronger the hybridization \cite{vanderLaan1986}. Assuming a realistic bandwidth instead of a single ligand state, these contributions are weighted and contribute much less than in the fm-CI model. Nevertheless van der Laan \textsl{et al.} concluded from their Anderson impurity calculation with 3\,eV broad bands and their x-ray absorption $M$-edge data that in intermediate valent compounds like CePd$_3$ the $J=7/2$ contribution may be as high as 30\,\% \cite{vanderLaan1986}. 

\begin{figure}[tbp]
   \includegraphics[width=1.00\columnwidth]{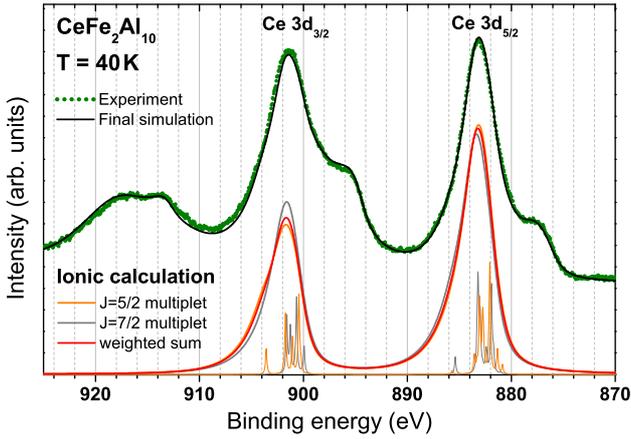}
    \caption{Ionic simulation of the Ce~$3d$ PES spectra of the pure $J=5/2$ (orange) and $J=7/2$ (gray) states. The calculations are shown both before (underlying multiplet structure scaled by $\times0.1$) and after the broadening functions have been applied. Here the same Gaussian, Lorentzian and Mahan broadening as in the simulations of the Ce$M_2$Al$_{10}$ HAXPES spectra (Fig.~\ref{plasmon_corr}) are used. The red curve is the weighted sum of 69\,\% $J=5/2$ and 31\,\% $J=7/2$. The experimental data of CeFe$_2$Al$_{10}$ (green dots) and the corresponding final simulation (black curve) are also shown (displaced on the y-axis by an offset).}
    \label{sevenhalf}
\end{figure}

For the $J=7/2$ contributions in the ground state our fm-CI calculations yield about 19\,\% in the case of Ru, 22\,\% for Os and approximately 31\,\% for the Fe compound. This shows nicely how the intermixing of the higher multiplet increases with increasing hybridization. Actually, the presence of a larger amount of $J=7/2$ in the Fe compound is also experimentally confirmed by the line shape of the Ce $M_5$ edge as measured with soft x-ray absorption in a previous experiment \cite{Strigari2013}. The low energy peak of the $M_5$ edge is expected to become stronger for larger contributions of $J=7/2$ in the ground state \cite{vanderLaan1986} which is in agreement with our observation (see Fig.\,3 in Ref.\,\cite{Strigari2013} or Fig.\,1(c) in Ref.\,\cite{Ishiga2014}). However, back to the HAXPES simulations, we know that in a full band model the intermixing with the higher multiplet would be weaker than the numbers above from the fm-CI calculations may suggest. Furthermore, we know that the line shape of the Ce~$3d$ $J=7/2$ emission multiplet differs from the $J=5/2$ one (see thin orange and gray lines in Fig.\,\ref{sevenhalf}). Hence the question arises whether the erroneous amount of $J=7/2$ influences the outcome of $f^0$ contribution in the initial state.

An estimation of this error is given by Fig.\,\ref{sevenhalf}, where the ionic simulations ($f^1$ only) of the Ce~$3d$ PES spectra of the pure $J=5/2$ (orange curves) and $J=7/2$ (gray curves) states are compared. Both the underlying multiplet structure and the broadened ionic spectra are shown; here the same broadening functions as in the final fm-CI simulations of the Ce$M_2$Al$_{10}$ HAXPES spectra are used. The red curve in Fig.\,\ref{sevenhalf} represents the weighted sum of the pure $J$ spectra, using 69\,\% of $J=5/2$ and 31\,\% of $J=7/2$. This corresponds to the mixing in the final fm-CI simulation for CeFe$_2$Al$_{10}$, which is shown in the same panel (shifted upwards by an offset) together with the experimental data. Note that in the final simulation shown in Fig.~\ref{plasmon_corr} the different $f^n$ contributions cannot be easily separated since their multiplet lines partly overlap. Thus, a direct extraction of the $f^1$-only contribution is not possible. Instead, consulting the ionic simulations allows to quantitatively compare the $J=5/2$ and $J=7/2$ line shapes.

Comparing the red and the orange curve in Fig.\,\ref{sevenhalf} shows that the presence of $J=7/2$ has only minor effects on the spectral line shape although the $J=7/2$-contribution is appreciable (31\,\%). The Ce~$3d_{5/2}$ feature ($880-890$\,eV) in the weighted sum (red) is nearly identical to the pure $J=5/2$ (orange). The differences in the Ce~$3d_{3/2}$ region are only slightly larger. At most, the intensity ratio between the spin-orbit split Ce~$3d_{5/2}\,f^1$ and $3d_{3/2}\,f^1$ multiplets is affected -- the possible error amounts to about $4$\,\%, meaning an error of less than $\pm 0.5$\,\% for the $f^0$ contribution. Here the systematic uncertainties due to background and plasmon corrections are more essential so that we conclude the error in the fm-CI model due to the overestimation of the $J=7/2$ multiplet in the ground state is small and in particular does not influence the result for $f^0$.

\bibliographystyle{model1a-num-names}

\end{document}